\documentstyle[preprint,aps]{revtex}
\begin{document}
\draft
\author{R. Prozorov, A. Poddar, E. Sheriff, A. Shaulov, Y. Yeshurun}
\address{Institute for Superconductivity, Physics Department, Bar-Ilan University,\\
52900 Ramat-Gan, Israel}
\title{Irreversible magnetization in thin $YBCO$ films rotated in external magnetic
field\thanks{{\it Accepted in Physica C}}}
\date{Received: (26-Feb-1996) Accepted: (17-April-1996)}
\maketitle

\begin{abstract}
The magnetization $M$ of a thin $YBa_2Cu_3O_{7-\delta }$ film is measured as
a function of the angle $\theta $ between the applied field $H$ and the $c-$
axis. For fields above the first critical field, but below the Bean's field
for first penetration $H^{*}$, $M(\theta )$ is symmetric with respect to $%
\theta =\pi $ and the magnetization curves for forward and backward rotation
coincide. For $H>H^{*}$ the curves are asymmetric and they do not coincide.
These phenomena have a simple explanation in the framework of the Bean
critical state model.
\end{abstract}


\section{Introduction}

The anisotropic properties of high-temperature superconductors have
motivated measurements of their magnetization $M$ as a function of the angle 
$\theta $ between the external DC magnetic field $H$ and one of the
principal axis of the superconductor. Such experiments may supply ample
information regarding the superconducting properties and they were used for
studies of the $3D$ anisotropic Ginzburg-Landau theory in single $%
YBa_2Cu_3O_{7-\delta }$ crystal \cite{pugnat}, pinning strength distribution
in polycrystalline materials \cite{mumtaz,goeckner,liu,khosa,hasan},
intrinsic anisotropy \cite{yaron,hasanain} and interaction of vortices with
pinning sites and external magnetic fields \cite{goeckner,hasan,kim}. The
conclusions in these works were drawn from the peculiarities observed in
rotation curves at different ambient conditions. Thus, for example, it was
found that in relatively high fields, well above the lower critical field $%
H_{c1}$, the $M\left( \theta \right) $ curves in forward and backward
rotations do not coincide \cite{goeckner,yaron,hasanain,kim}.This was taken
as an evidence that the magnetic moment rotates frictionally, lagging behind
the sample, due to the interaction of vortices with the external field, thus
exhibiting irreversible (hysteretic) behavior and an apparent phase shift
between the two magnetization curves. Such a phase shift was not observed in
lower fields, where the two curves were found to coincide and it has been
believed to indicate that the sample is in a pure reversible (or Meissner)
state. The main purpose of this work is to demonstrate that the forward and
backward rotation curves coincide even in the irreversible state, for fields
between $H_{c1}$and $H^{*}$, and that the observed hysteretic behavior in
higher fields may find a simple explanation within the framework of the Bean
critical state model \cite{bean}.

We present here measurements of the angular dependence of the magnetization
curves $M(\theta )$ of a thin $YBa_2Cu_3O_{7-\delta }$ (YBCO) film. As
demonstrated below, in thin films we can safely neglect the in-plane
component of magnetic moment. Therefore, the results of rotational
experiments are clear and much easier to interpret. We find that for fields 
{\em below} the Bean's first penetration field $H^{*}$, $M(\theta )$ is
symmetric with respect to $\theta =\pi $ and the rotation curves for forward
and backward rotation coincide (see also \cite{yaron}). For $H>H^{*}$ the
curves are asymmetric and they do not coincide. Moreover, the backward curve
is a mirror image (with respect to $\theta =\pi )$ of the forward one.

The basic idea of our explanation is that in order to understand the
variation of $M$ during rotation in a system with isotropic pinning one has
to consider separately the variation of its components along the $c-$ axis ($%
M_c$) and the $ab-$ plane ($M_{ab}$). This leads to a consideration of the
projections of the applied field on the $c-$ axis ($H_c$) and the $ab-$
plane ($H_{ab}$), respectively. The effective field $H_c$ cycles during the
rotation between $\pm H$ and the calculation of $M\left( \theta \right) $ is
thus analogous to the calculation of $M\left( H\right) $. We demonstrate the
validity of this concept by comparing direct measurements of $M\left(
H\right) $ and $M\left( \theta \right) $. We stress, that this claim is true
only if a sample does not have any induced anisotropy of pinning, e.g., twin
boundaries or columnar defects. Although, the analysis below can be easily
extended to account for more general, anisotropic case.

Many authors pointed out the importance of the geometry for proper analysis
of the rotation experiment. In previous works this was limited to a
consideration of the demagnetization in a reversible state \cite
{yaron,kolesnik}, i.e. to the actual field on the sample edge which varies
during rotation because of demagnetization. Maintaining this point, we
further show how demagnetization affects the rotation curves in an
irreversible state. In particular, we find that the interval of angles for
which the sample undergoes remagnetization shrinks dramatically as a result
of a flat geometry. This results in sharp changes in $M\left( \theta \right)
.$

\section{Experimental}

A thin {\it YBCO} film of thickness $\simeq 1000$ \AA\ and lateral
dimensions of $5\times 5$ $mm^2$ was laser ablated on a $SrTiO_3$ substrate 
\cite{koren}. The film is $c-$ oriented, so that the $c-$ axis points in a
normal to a film surface direction with angular dispersion less than $2^o$
and the $ab-$ plane coincides with the film plane. For the rotation magnetic
measurements we used an ''{\it Oxford Instruments'' Vibrating Sample
Magnetometer (VSM)} that enables sample rotation relative to an external
magnetic field with a $1^o$ precision. The rotation axis is always
perpendicular to the $c-axis$. Samples were zero field cooled down to the
desired temperature whence an external magnetic field $H$ was turned on. The
component of the magnetic moment $M_H$ along the external field direction
was then measured{\it \ }while the sample was rotated one full turn
(''forward rotation'') and then back (''backward rotation''). The initial
field application was always along the $c-$axis. We find that turning on the
external field at other angles does not yield any new information, since the
rotation curve becomes independent of this angle after one full rotation -
(see also \cite{hasanain}).

Throughout this paper $M_H$ is the magnetic moment component along the
direction of the external field $H$. We recall that in a VSM, as well as in
most other magnetometers, $M_H$ is the {\em measured} component of the total
magnetic moment. In Fig. \ref{geometry} we sketch the relevant vectors and
angles.

\section{Results}

The symbols in Fig. \ref{partialfilm} present the measured angular
dependence of the zero-field-cooled (ZFC) magnetization, $M_H(\theta )$, for
the {\it YBCO} film at different values of $H$ and at $20\ K$. (At this
temperature, the apparent $H_{c1}\approx 50$ $G$, and $H^{*}=380$ $G$ , as
determined from direct ''static'' $M(H)$ measurements). The rotation curves
are symmetric with respect to $\theta =\pi $, in spite of the fact that the
applied fields are larger than $H_{c1}$. Also, we find that the
magnetization curves for forward and backward rotation coincide. This
implies that such reversibility with respect to direction of rotation is not
indicative of the ${\em true}$ magnetic reversibility. Note, that curve for $%
H=400\ G>H^{*}$ is shown only for comparison.

Figures \ref{abovehst}-\ref{asymmetry} exhibit $M_H(\theta )$ data collected
at $20\ K$ for fields above $H^{*}$. Fig. \ref{abovehst} demonstrates that
as a result of an increase in the external field, the rotation curves become
gradually more and more asymmetrical with respect to $\theta =\pi .$ For
field slightly larger than $H^{*}$ the forward and backward rotation curves
look harmonically, but with some phase shift, see Fig. \ref{phaseshift}.
High - field measurements, shown in Fig. \ref{asymmetry} demonstrate that
the backward rotation curve is a mirror image of the forward one with
respect to $\theta =\pi $. Therefore, we conclude that the observed
asymmetry do not imply a phase shift, but a true magnetic hysteresis with
respect to rotation, which means a reverse of the magnetic moment when the
direction of rotation is reversed. We shell return to this point in the
analysis.

We complete the experimental picture by presenting in Fig. \ref{fullfilm}
the rotation curves at $H=1.5$ $Tesla$ measured at different temperatures.
The width and height of steep change in a magnetic moment shrinks as the
temperature increases and, in some sense, the increase of temperature is
analogous to the increase of magnetic field. As we show below, all these
features find natural explanation in a framework of the Bean model.

\section{Analysis}

We present a model that takes into account the variation of the {\em %
effective }field along the sample sides during rotation. Specifically,
during a rotation cycle, $H_c,$ the field component along the $c-$axis,
oscillates as $H_c=H\cos \left( \theta \right) $. This leads to a variation
of $M_c$ similar to that in a standard magnetization loop, where the
magnetization $M_c$ is measured as a function of the external magnetic field
between $-H$ to $+H$ at $\theta =0$. Therefore, in order to analyze a
rotation experiment one has to consider the {\em effective} applied field $%
H_c$ along the $c$ - axis, and not the actual applied field $H.$ This
approach leads directly to the consideration of the two field regimes: ($a$) 
\underline{{\sl Moderate fields:}} $H\leq H^{*}$, where the sample space is
only partially occupied by magnetic flux and ($b$) \underline{{\sl High
fields}:} $H\geq H^{*}$, where flux occupies the entire sample space. These
two regimes are discussed below. We note that for fields below $H_{c1}$ the
correct analyses of the data was given previously in a number of reports,
e.g. \cite{yaron,kolesnik}.

In order to elucidate the relative importance of each component of the
applied field we start by calculating the components of the magnetization
along the $c-$ axis, $M_c$, and in the $ab-$ plane, $M_{ab}$. For an
estimate of the relative contribution of $M_c$ and $M_{ab}$ to the total
magnetic moment we apply the Bean model to a finite slab as, for example, in
Refs. \cite{bean,hellman,yeshurun}. (The validity of this approach was
examined experimentally on both films \cite{hellman} and crystals \cite
{yeshurun}). Then (by taking $t\leq L\leq d$, see Fig. \ref{geometry}) one
obtains:

\[
\left\{ 
\begin{tabular}{l}
$\left| M_c\right| \approx \frac{\left| J_c^a\right| }{40}L\left( 1-\frac L{%
3d}\right) \equiv \frac{H_c^{*}}2$ \\ 
$\left| M_{ab}\right| \approx \frac{\left| J_c^c\right| }{40}t\left( 1-\frac 
t{3L}\right) \equiv \frac{H_{ab}^{*}}2$%
\end{tabular}
\right. \Longrightarrow 
\]

\begin{equation}
\beta \equiv \left| \frac{M_c}{M_{ab}}\right| \approx \frac{H_c^{*}}{%
H_{ab}^{*}}=\frac{\left| J_c^{ab}\right| }{\left| J_c^c\right| }\frac Lt%
\frac{\left( 1-\frac L{3d}\right) }{\left( 1-\frac t{3L}\right) }
\label{aleph}
\end{equation}

\noindent where the magnetization is in $emu/cc$, current densities are in $%
A/cm^2$ and lengths are in $cm.$ $H_c^{*}$ and $H_{ab}^{*}$ are the
effective penetration fields along the $c$-axis and the $ab$-plane,
respectively. $J_c^{ab}\left( H_c\right) $ and $J_c^c\left( H_{ab}\right) $
are the persistent current densities flowing in and out of the $ab$-plane,
respectively. For our sample (typical for thin films) $d=L=0.5$ {\it cm}, $%
t=10^{-5}$ {\it cm}, the above ratio becomes approximately $\beta \approx
3\cdot 10^4\frac{\left| J_c^{ab}\right| }{\left| J_c^c\right| }.$ This shows
that in the case of thin film we can safely omit the in-plane component of
the magnetic moment $M_{ab}$ and in-plane component of the applied field - $%
H_{ab}$. We note that this simplification is not crucial for the analysis.
Moreover, one ought to include both components of magnetic moment and field
analyzing data for thick samples. This can be easily done using the same
approach, as we undertake below.

As it is stated above, the {\em measured} magnetization, $M_H$, in a VSM, as
well as in many other techniques, is the component of the magnetization
along the external field. It is convenient to express $M_H$ as $M_H\left(
\theta \right) =\left| \overrightarrow{M\left( \theta \right) }\right| \cos
\theta =M_c\left( \theta \right) \cos \theta +M_{ab}\left( \theta \right)
\sin \theta $. For thin films, as shown above, one may safely rewrite this
equation as $M_H\left( \theta \right) =M_c\left( \theta \right) \cos \theta $%
. Below we use $H^{*}=H_c^{*}$.

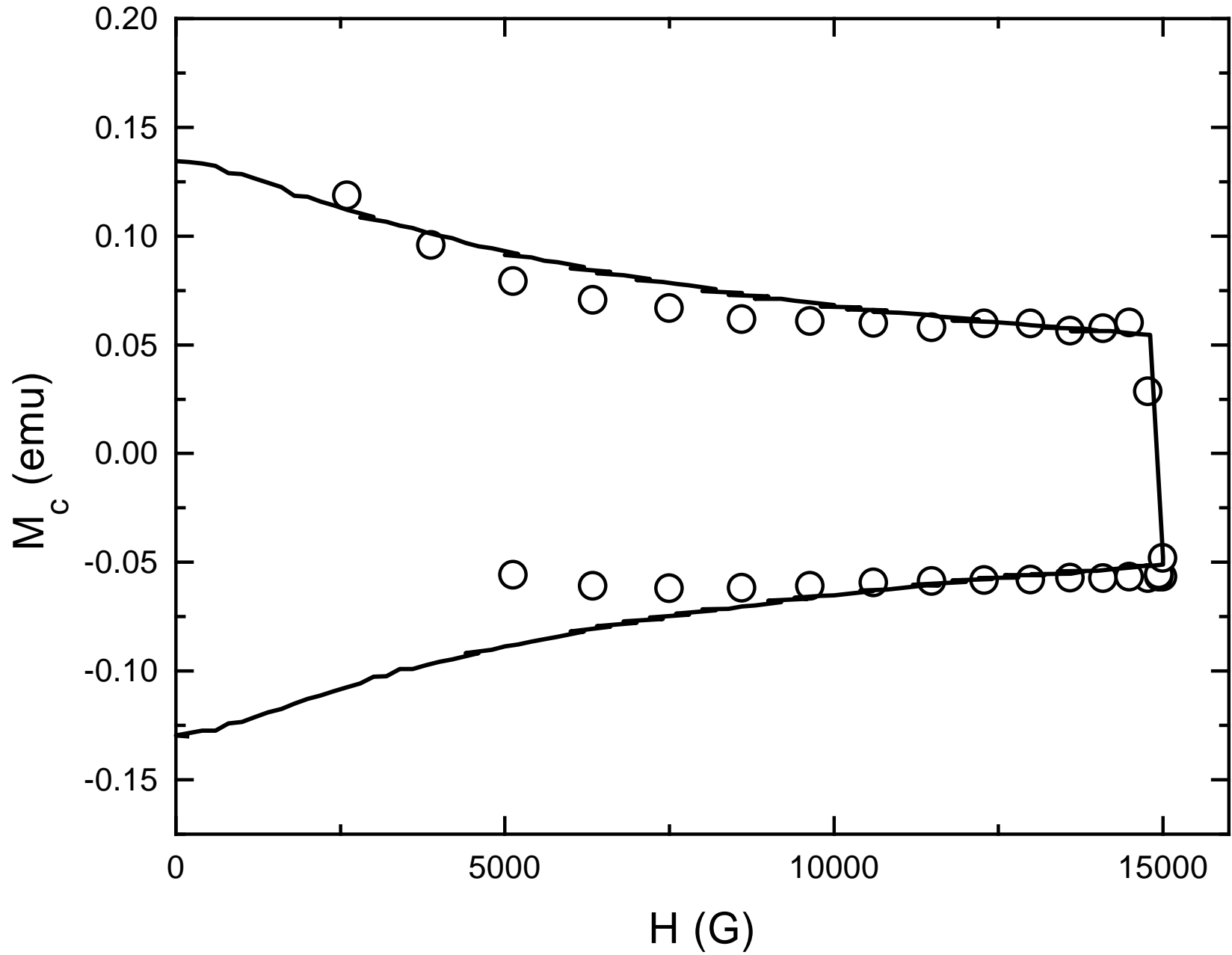
\begin{figure}[tbh]
\caption{Comparison of the standard magnetization loop $M\left( H\right) $
(solid line) with the loop, constructed from the rotation experiment (open
circles) as described in the text.}
\label{m(h)vsm(t)}
\end{figure}

It should be noted, that for a sake of clarity, the analysis below is based
on the Bean model for an infinite slab. Whereas in a fully magnetized state
the magnetization for a thin sample and for an infinite slab is the same and
given by Eq. \ref{aleph}, the remagnetization process is quite different 
\cite{zeldov}. Nevertheless, we use the simple Bean model first, in order to
demonstrate a general approach to the problem avoiding an unnecessary
complications of the analysis. As we show latter (Fig. \ref{m(h)vsm(t)}) we
could even use a linear approximation for a remagnetization stage.

\subsubsection*{Partial magnetization ($H\leq H^{*}$)}

Utilizing the parameter $x=1-\cos \left( \theta \right) $ one may express
the difference between the external magnetic field $H$ and its projection on
the $c$-axis during rotation, as $\Delta H=Hx$. In the following we describe
a ZFC experiment and consider {\em forward} rotation only. The {\em backward}
rotation may be obtained from the formulae below by substituting $\theta
_{back}=2\pi -\theta $. The curve obtained by such a substitution {\em %
coincides} with the forward curve.

In the framework of the Bean model we get for the projection of the total
moment along the $c$-axis:

\begin{equation}
M_c=-\frac{H^2}{8H^{*}}\left( x^2+2x-4\right) -H\left( 1-x\right) .
\label{partialmc}
\end{equation}

Note that for $x=0$ we recover the Bean result for partial magnetization.
The component $M_c$ varies continuously with $\theta $ in a whole interval
of angles implying that the magnetic flux profile inside the sample changes
for any change in $\theta $. In the following we refer as 'remagnetization'
to the parts of the process for which the profile is changing.

Eq. \ref{partialmc}, when expressed in terms of $\theta $, yields, for the
measured component of the total moment along the direction of the external
magnetic field:

\begin{equation}
M_H=H\cos \left( \theta \right) \left\{ \frac H{8H^{*}}\sin ^2\left( \theta
\right) -\left( 1-\frac H{2H^{*}}\right) \cos \left( \theta \right) \right\}
.  \label{partialmh}
\end{equation}
Apparently, the magnetization curves for {\em backward} and {\em forward}
rotations are symmetric with respect to $\theta =\pi $ and therefore they 
{\em coincide}. In other words, reversibility with respect to the direction
of rotation does not imply a ''{\sl true}'' magnetic reversibility which is
expected either in the Meissner state or in the unpinned state.

The magnetic moment along the $c$-axis reaches a maximum value of 
\begin{equation}
\left| M_c\right| _{\max }=\left| M_H\right| _{\max }=H_{ab}\left( 1-\frac H{%
2H^{*}}\right)  \label{partialmcmax}
\end{equation}

\noindent at $x=0$ and $2$, $\left( \text{i.e. }\theta =0,\pi \text{ and }%
2\pi \right) $.

Fig. \ref{partialfilm} shows a good agreement between Eq. \ref{partialmh}
and the experiment. In this figure the symbols represent the experimental
data whereas the solid lines are fits to Eq. \ref{partialmh} with a single
parameter $H^{*}\approx 380$ {\it G }, {\em for all curves. } The value of $%
H^{*}$ was determined from a fit of the maximum value of $M_H$ to Eq. \ref
{partialmcmax} (inset to Fig. \ref{partialfilm}) and was verified through
independent measurements of standard magnetization loops in that sample.

Another implication of Eq. \ref{partialmh} is that as long as the applied
field $H$ is smaller than or equal to $H^{*},$ the component $M_H$ of the
total magnetic moment is less than (or equal to) zero in the whole angular
range. We show below that for $H>$ $H^{*},$ $M_H$ becomes positive at
certain angles. This crossover from negative to positive values of $M_H$ may
serve as a sensitive tool for experimental determination of $H^{*}$. In Fig. 
\ref{partialfilm} this crossover occurs at $H=H^{*}\approx 380$ $G${\it . }%
An additional line at $H=400$ $G>H^{*}$ is shown for comparison.

\subsubsection*{Full magnetization ($H\geq H^{*}$)}

When the applied field is larger than $H^{*}$ magnetic flux penetrates the
entire sample space. In this case, the projection of the magnetic moment
along the $c$-axis in the interval $x=\left[ 0,2\right] $ $($i.e. $\theta
=\left[ 0,\pi \right] )$ according to the Bean model is:

\begin{equation}
M_c=\left\{ 
\begin{array}{lc}
\frac 34Hx-\frac 18\frac{H^2}{H^{*}}x^2-\frac{H^{*}}2 & \quad x\leq 2\frac{%
H^{*}}H \\ 
\frac{H^{*}}2 & \quad x\geq 2\frac{H^{*}}H
\end{array}
\right.  \label{mcfull}
\end{equation}

\noindent Again, we note that for $x=0$ we recover the Bean results for full
penetration $\left| M_c\right| =H^{*}/2$. For $x\geq 2\frac{H^{*}}H$ the
magnetization is constant as predicted by Bean for $H>H^{*}$. Only for $%
x\leq 2\frac{H^{*}}H$ we get a non trivial result which reflects the fact
that the effective field is being reversed. Thus, the remagnetization
process is limited now to $x\leq 2\frac{H^{*}}H$ and it is completed when
the moment reverses its sign (i.e. changes from $-\frac{H^{*}}2$ to $+\frac{%
H^{*}}2)$.

The measured component of the magnetic moment along the direction of the
external field may be determined from Eq. \ref{mcfull} by substituting $%
x=1-\cos \left( \theta \right) $:

\begin{equation}
M_H=\left\{ 
\begin{array}{lc}
\frac H8\cos \left( \theta \right) \left( -\frac H{H^{*}}\left( 1+\cos
\left( \theta \right) ^2\right) +2\left( \frac H{H^{*}}-3\right) \cos \left(
\theta \right) +6-\frac{4H^{*}}H\right) & ~\theta \leq \theta _r \\ 
\frac{H^{*}}2\cos \left( \theta \right) & ~\theta \geq \theta _r
\end{array}
\right.  \label{mhfull}
\end{equation}

\noindent where, $\theta _r=\arccos \left( 1-2H^{*}/H\right) \leq \pi $ is
the angle at which the remagnetization process is completed.

An interesting implication of Eq. \ref{mhfull} is that for $H>H^{*}$ the
resulting magnetization versus angle curves become {\em asymmetric} with
respect to $\theta =\pi $. Hence, the {\em backward} rotation curve {\em %
does not coincide} with the {\em forward} rotation curve. We thus assert
that generally, a {\em forward} and a {\em backward} magnetization versus
angle curves would not coincide if $H>H^{*}$ and not, as previously
believed, when $H>H_{{c1}}$. Also, the backward curve is a mirror image of a
forward one with respect to $\theta =\pi $.

The solid lines in Fig. \ref{fullfilm} are fits to Eq.\ref{mhfull}. The
sharp change in $M_H$ indicates a reversal of the magnetic moment $\left(
\Delta M=H^{*}\right) $ within a narrow angular interval. We explain this
feature in the next section by considering the demagnetization effects.

\subsubsection*{Demagnetization effects}

One may regard demagnetization effects as a renormalization of the applied
magnetic field. In fact, to be more precise, the applied field $H$ in the
above formulae should be replaced with the actual magnetic field intensity
at the specimen edges, which is in the simplest form: $H^{eff}=H+\gamma
H^{*} $, were $\gamma $ is a dimensionless parameter accounting for
demagnetization. (Note the difference with the usual notion for a
demagnetization correction for reversible state).

The remagnetization region, as described above, occurs when the projection
of the applied magnetic field $H_c$ changes sign. In standard magnetization
- loop measurements it happens twice in each full loop and it was analyzed
previously, see, e.g., \cite{bean,brandt1}. One may therefore refer to the
analysis of a standard magnetization measurements in order to gain
understanding with regard to its effects within rotation experiments. The
magnetic field interval, within which the remagnetization occurs for
infinite slab is $\Delta H=2H^{*}$ for infinite slab. However, for finite
sample, one must bear in mind that when referring to this interval one
actually refers to the {\em effective} magnetic field on the sample edges $%
\Delta H^{eff}=2H^{*}=\Delta H+\gamma H^{*}$, or $\Delta H=\left( 2-\gamma
\right) H^{*}$. Thus, since the demagnetization effects lead to an increase
of $H^{eff}$ with a decrease of sample thickness, the applied external field
interval for remagnetization shrinks for thinner samples \cite{oussena}.
Demagnetization effects therefore change the angular interval for
remagnetization to complete: $\theta _r^{eff}=\arccos \left(
1-2H^{*}/H^{eff}\right) $. The total change in magnetic moment retains its
original value $\Delta M=H^{*}$. This conclusion is in a good agreement with
presented data on thin film, where we find a very narrow angular interval
for remagnetization, within which the moment of a large magnitude changes
sign.

Finally, in order to verify experimentally our assumption about the
similarity between the standard magnetization loops and the rotation
experiment we show in Fig. \ref{m(h)vsm(t)} a standard $M\left( H\right) $
loop (solid line) measured with the applied field along the $c$ - axis,
along with a ''converted'' $M_c\left( H_c\right) $ loop, i.e. $M_c=M_H/\cos
\left( \theta \right) $ versus $H_c=H\cos \left( \theta \right) $. The
remarkable similarity between the two curves supports our approach in
explaining the data.

As we noted above, such correspondence of loops is possible only, if a
sample does not have induced extrinsic anisotropy. In the case of
anisotropic pinning, in a first approximation in the expressions above, the
characteristic penetration field should be replaced by $H^{*}\left( \theta
\right) $. More rigorous treatment requires exact analysis of the magnetic
flux evolution in the sample during rotation with subsequent calculation of $%
M\left( \theta \right) $.

\section{Summary and conclusions}

Detailed analysis based on the Bean model of the irreversible magnetization
of a rotating type-II superconductor is presented. We assert that during
rotation, the magnetic moment changes its sign with respect to the $c-$axis.
This remagnetization happens within a finite angular interval, yielding an
asymmetric rotation curve. All main features observed in the experiment are
explained from this point of view. It is shown that demagnetization does not
change the functional dependence of $M_H$ vs. angle curve, however it does
affect it by shrinking the angular interval within which the remagnetization
occurs.

{\sl Acknowledgments: }We thank Gad Koren for providing us with $YBCO$ thin
films. This research was supported in part by the Israeli Science Foundation
administered by the Israeli Academy of Science and Humanities and, in part,
by the Heinrich Hertz Minerva Center for High Temperature Superconductivity.
Y. Y. acknowelgedges a support from the DG XII, Commission of the European
Communities and the Israeli Ministry of Science and the Arts (MOSA). A. S.
acknowledges a support from the Israeli-France cooperation program AFIRST.

\begin{figure}[bh]
\caption{Geometrical aspects of the experiment. The rotation axis is always
perpendicular to the $c-$axis.}
\label{geometry}
\end{figure}

\begin{figure}[tbh]
\caption{Angular dependence of the ZFC magnetization ($M_H$) at 20 K for 
{\it YBCO} film at different values of $H$. Symbols are experimental points
and solid lines are calculated from Eq. \ref{partialmh}. \protect\underline{%
{\it Inset}} shows the variation of maximum moment $\left| M_H\right| _{\max
}$ with $H$. Solid line is a fit to Eq. \ref{partialmcmax}.}
\label{partialfilm}
\end{figure}

\begin{figure}[tbh]
\caption{Angular dependence of the ZFC magnetization ($M_H$) at 20 K for 
{\it YBCO} film at different values of $H>H^{*}$. }
\label{abovehst}
\end{figure}

\begin{figure}[tbh]
\caption{Forward and backward rotation curves measured at 20 K for {\it YBCO}
film at $H=400\ G$. }
\label{phaseshift}
\end{figure}

\begin{figure}[tbh]
\caption{Forward and backward rotation curves measured at 20 K for {\it YBCO}
film at $H=15000\ G$. }
\label{asymmetry}
\end{figure}
\begin{figure}[tbh]
\caption{Angular dependence of the magnetisation $M_H$ at different
temperatures for {\it YBCO} thin film, $H=15000\ G$. }
\label{fullfilm}
\end{figure}

\end{document}